\documentclass[10pt]{article}
\usepackage[utf8]{inputenc}
\usepackage{amsmath}
\usepackage{bbold}
\usepackage[a4paper, margin=2cm]{geometry}
\usepackage[english]{babel}
\usepackage{graphicx}
\usepackage{float}
\usepackage{fontspec} 
\usepackage{unicode-math}
\usepackage{etoolbox}
\usepackage{wasysym}
\usepackage{gensymb}
\usepackage{url}
\usepackage{cite}
\patchcmd{\thebibliography}{\section*{\refname}}{}{}{}

\usepackage{wrapfig}
\usepackage[table,xcdraw]{xcolor}

\usepackage{siunitx}
\usepackage{titlesec}

\usepackage{titlesec}

\titleformat{\paragraph}
{\normalfont\normalsize\bfseries}{\theparagraph}{1em}{}
\titlespacing*{\paragraph}
{0pt}{3.25ex plus 1ex minus .2ex}{1.5ex plus .2ex}

\date{\vspace{-10ex}}

\begin{document}
\title{Spatial resolution of X-ray beam-tracking microscopy \\
}
\maketitle

\begin{center}
    \textbf{{\large }}
    \\
    {\large\textbf{Harry Allan}$^{1,^*}$}, {\large Carlos Navarrete-León$^{1}$}, {\large Adam Doherty$^{1}$}, {\large Shashidhara Marathe$^2$}, {\large Kaz Wanelik$^2$}, 
    \large{Marco Endrizzi$^{1}$}
    \\
    \textit{$^1$Department of Medical Physics and Biomedical Engineering, University College London, London, UK} \\
    \textit{$^2$Diamond Light Source, Harwell Science and Innovation Campus, Didcot, UK} \\

\end{center}

$^*$ Correspondence email: harry.allan.21@ucl.ac.uk
\begin{abstract}

X-ray beam-tracking is a phase-contrast imaging technique capable of simultaneously retrieving transmission, phase, and dark-field images. Although the spatial resolution in beam-tracking is largely considered to be 'aperture driven', no model yet exists to describe this in full. The dark-field channel is of particular interest, due to previous observations of anomalously high sharpness compared to transmission and phase channels. We derive a full optical transfer function model for each contrast channel using the Fokker-Planck equation for near-field imaging. Experimental validation using both synchrotron-based and laboratory-based setups, with 15 \textmu{}m circular and 10 \textmu{}m rectangular apertures, reveals a limiting resolution of at least 3 \textmu{}m, much smaller than the apertures themselves. Together, the model and the supporting experiments offer a full description of spatial resolution in beam-tracking, and formally confirm the greater spatial resolution in the dark-field channel. These findings open new possibilities in system design and experimental protocols to exploit these capabilities.

\end{abstract}
\section{Introduction}

\begin{figure}
    \centering
    \includegraphics[width=0.99\linewidth]{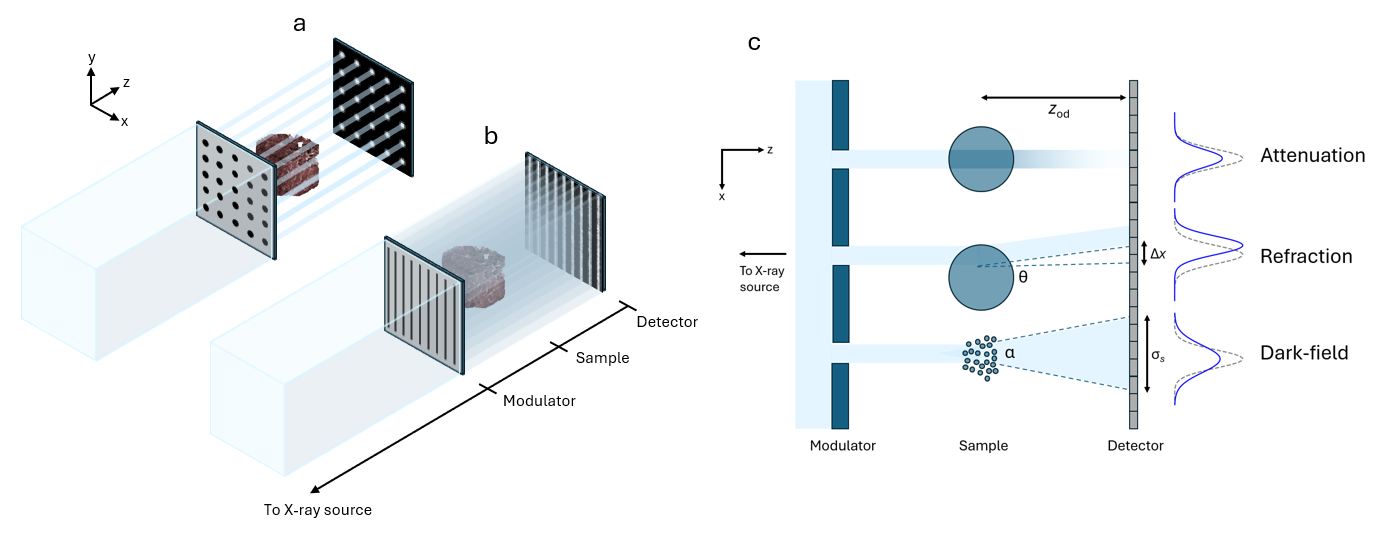}
    \caption{General geometries of X-ray beam-tracking imaging systems, utilising either 2D (a) or 1D (b) apertures. The attenuating modulator shapes the beam into an array of spatially separated beamlets, which then impinge upon the sample. The sample-induced perturbations are measured after some propagation distance $z_\text{od}$, by directly resolving them with the detector (c). Attenuation results in a reduction of the beamlet intensity, refraction results in a translation of the beamlet centroid, and dark-field results in an increase in the beamlet width.}
    \label{fig:system_diagrams}
\end{figure}

X-ray phase-contrast imaging is a broad class of techniques deriving contrast from sample-induced phase shifts upon an X-ray beam \cite{wilkins2014evolution,endrizzi2018x}. This enables the imaging of samples that have no or limited attenuation contrast, and can result in large reductions in dose requirements \cite{kitchen2017ct}. Among a number of other techniques \cite{wilkins1996phase,weitkamp2005x,pfeiffer2008hard,olivo2001innovative,olivo2007coded,wen2008spatial,morgan2011quantitative,morgan2012x,berujon2012two,gustschin2021high}, X-ray beam-tracking \cite{vittoria2015beam} is promising for its simple setup, relaxed alignment requirements, and the ability to access multi-contrast information with a single acquired image. The principle of the technique is that an attenuating modulator is used to spatially structure the X-ray beam into an array of independent beamlets, from which sample-induced perturbations may be measured and quantified \cite{wilkins1995improved}. Each beamlet acts as an independent, self-contained, imaging system, thus beam-tracking could be likened to a highly parallelised version of scanning transmission x-ray microscopy \cite{kaulich2002diffracting,de2008quantitative,takeuchi2013three}. The structured illumination array may be arranged as either 1D line apertures \cite{vittoria2015beam, brombal2023pepi} (figure \ref{fig:system_diagrams}b), or as 2D apertures \cite{dreier2020tracking,navarrete2023x,lioliou2023laboratory,zhou2025high,li2025beam} (Figure \ref{fig:system_diagrams}a), allowing the extraction of additional directional information \cite{dreier2020tracking,navarrete2025laboratory}. Assuming ample beamlet independence and depth-of-focus, Shack-Hartmann like configurations, using lenslet arrays or other optical-elements  \cite{mayo2004refractive,reich2018scalable,mikhaylov2020shack,mamyrbayev2020development}, can also be described by the beam-tracking framework. The versatility of the approach has enabled implementation in both compact laboratory systems \cite{navarrete2023x}, and as a highly sensitive synchrotron setup \cite{navarrete2024high} for imaging of challenging, low-contrast samples.
\par
In order to track the perturbations of the beamlets, the experimental apparatus is arranged such that each beamlet is sampled by a number of detector pixels. This is achieved through magnifying geometries \cite{vittoria2015beam,navarrete2023x}, high-resolution detectors \cite{vittoria2017multimodal,esposito2022laboratory}, or sub-pixel localisation using photon counting detectors \cite{dreier2020tracking}. Attenuation results in a simple reduction of the beamlet intensity. By introducing the propagation distance $z_\text{od}$ after the sample, refraction of angle $\theta$ results in a translation (in the small angle approximation) of the beamlet centroid by a distance $\Delta x = \theta z_\text{od}$. The dark-field signal, arising predominantly from diffuse scattering by microstructure below the aperture width, induces a fan of forward-scattered photons with opening angle $\alpha$. Similarly, propagation causes a broadening of the beamlet up to a width $\sigma_s = \text{sqrt}( \alpha^2 z^2_\text{od} + \sigma^2 )$, where $\sigma$ is the width of the unperturbed beamlet. Each of these beamlet perturbations are illustrated in Figure \ref{fig:system_diagrams}c.

While beam-tracking shares similarities with a number of other tracking-based methods \cite{wen2008spatial,morgan2011quantitative,morgan2012x,berujon2012two,gustschin2021high}, it differs in the fact that the small open fractions, low duty-cycles, and near-opaque modulators, generate highly independent beamlets, which sample specific localised regions of the sample. The independent beamlets limit the area of the beam that can contribute to image formation, leading to a spatial resolution different to that defined solely by the characteristics of the X-ray source or detector. When applied as a single-shot technique, the spatial resolution is sampling limited by the period of the illumination. More often, the spatial resolution is improved by combining multiple acquisitions while translating either the modulator or the sample, in a process known as dithering. The image formation process in beam-tracking is very similar to edge-illumination \cite{olivo2007coded,vittoria2014virtual}, the key difference being the use of a second 'detector mask' as an analyser, enabling larger detector pixels to be used. In this context, a number of previous works \cite{vittoria2015beam,dreier2020tracking,massimi2021dynamic,esposito2022laboratory,lioliou2023laboratory,doherty2023femtosecond,allan2026multi} have assumed the direct application of a model for spatial resolution in edge-illumination \cite{diemoz2014spatial} to also describe that for beam-tracking. 
\par
In this work, we derive a full analytical optical transfer function (OTF) model for the spatial resolution of X-ray beam-tracking. We derive separate expressions for the transmission and phase-contrast channels, demonstrating how they differ from edge-illumination. Crucially, we derive an expression for the dark-field contrast channel. This confirms previous observations of enhanced sharpness in dark-field imaging \cite{esposito2022laboratory,doherty2023edge,doherty2024hybrid}. The theoretical results in this work are supported by quantitative experimental data collected using 2D beam-tracking at a synchrotron setup, and 1D beam-tracking in a lab-based setup, demonstrating spatial resolution on the order of micrometres.

\section{Theoretical analysis}

\label{sec:theory}

We begin our analysis from the X-ray Fokker-Planck equation \cite{paganin2019x,morgan2019applying}, which describes the evolution of the intensity $I(x,y,z)$ of a paraxial complex coherent scalar wave field along the optical axis $z$

\begin{equation}
    \frac{\partial I(x,y,z)}{\partial z} = 
    -\frac{1}{k} \nabla_\perp \cdot \left[
    I(x,y,z)\nabla_\perp\phi(x,y,z)
    \right]
    + \nabla_\perp^2 \left[ D(x,y,z) I(x,y,z) \right].
    \label{FP_PDE}
\end{equation}

where $k = 2\pi/\lambda$ is the wavenumber, $\phi(x,y,z)$ is the phase of the wave, and $D(x,y,z)$ is a function describing signal diffusion due to scattering by sample microstructure. In near-field conditions with propagation distance $z_\text{od} \geq 0$ and Fresnel number $N_F \gg 1$, we may approximate the propagated intensity $I_z(x,y)$ using the first order Taylor expansion

\begin{align}
\begin{split}
I_z(x,y) \approx& \, T_0(x,y) I_\text{in}(x,y) -\frac{z_\text{od}}{k} \nabla_\perp \cdot 
   \left[ T_0(x,y) I_\text{in}(x,y)\nabla_\perp\phi_0(x,y) \right] \\
& \qquad\qquad\qquad\qquad + z_\text{od}\nabla_\perp^2 
   \left[T_0(x,y) I_\text{in}(x,y) D_0(x,y)  \right],
   \label{FP_FD}
\end{split}
\end{align}

where $I_\text{in}(x,y)$ is the intensity at the sample plane before interacting with the sample. In the projection approximation, $T_0(x,y)$ is the transmission of the wave-field through the sample, $\phi_0(x,y)$ is its phase shift, and $D_0(x,y)$ is the position-dependent diffuse scattering coefficient.
\par
We can now use Equation \ref{FP_FD} to model the intensity evolution $I(x,y,z)$ of a single beamlet. Assuming that the structured illumination period $p \gg \text{sqrt}( \sigma^2_D + \sigma^2_M )$, where $\sigma_D$ is the standard deviation of the detector point spread function (PSF), and $\sigma_M$ is the magnified width of the X-ray source spot projected onto the detector plane, neighbouring beamlets may be treated as independent. This allows us to model a single beamlet based on its normalised statistical moments $M_{n}[I(x,y)]$, where n is the order of the moment \cite{peiffer2023equivalence,buchanan2023direct}. 
\par
To analyse the system spatial resolution, we will calculate the response for each contrast channel to a 1D unit impulse. We model a dithered data acquisition strategy, which forms images in each contrast channel at arbitrarily finely sampled points $(m,n)$. The retrieved image value $A(m,n)$ (where $A$ is transmission, refraction, or dark-field) is thus the response of the system when the sample is at position $S(x-m,y-n)$ relative to the input beamlet $I_\text{in}(x,y)$. This framework allows the calculation of contrast-specific line spread functions (LSFs), from which OTFs may be derived.
\par
Image retrieval requires knowledge of both $I_z(x,y)$ and the input $I_\text{in}(x,y)$. In practice, it is possible to estimate $I_\text{in}(x,y)$ through a reference measurement of the propagated beamlet in the absence of the sample. In the region of validity of Equation \ref{FP_FD}, the reference beamlet intensity without the sample reduces to $I_\text{in}(x,y)$, which can be demonstrated by inserting $\nabla_\perp\phi_0(x,y)$ = $\nabla_\perp^2 D_0(x,y)$ = 0 and $T_0(x,y) = 1$ into Equation \ref{FP_FD}. The following derivations will thus assume that $I_\text{in}(x,y)$ is known. 

\subsection{Transmission}

Consider a pure attenuation signal by setting $\nabla_\perp\phi_0(x,y)$ = $ D_0(x,y)$ = 0 (or setting $z_\text{od} = 0$). Inserting this into Equation \ref{FP_FD}, we obtain $I_z(x,y) = T_0(x,y)I_\text{in}(x,y)$. The retrieved transmission at object coordinate $(m,n)$ is thus described by

\begin{align}
\begin{split}
    T(m,n) &= \frac{M_0[I_z(x,y)]}{M_0[I_\text{in}(x,y)]}
    \\
    &= \frac{\int^w_{-w}\int^w_{-w} T_0(x,y)I_\text{in}(x,y) \,dy\,dx}{\int^w_{-w}\int^w_{-w} I_\text{in}(x,y) \,dy\,dx},
    \label{eq:transmission_moments}
\end{split}
\end{align}

where the integral limits span the analysis window of half-width $w$, corresponding to a unit cell around the beamlet. To compute the LSF, we analyse the response of the transmission to a 1D unit line impulse $T_0(x,y) = \delta(x-m)$. Inserting the impulse into Equation \ref{eq:transmission_moments}, we show the LSF as a function of $m$

\begin{align}
    \text{LSF}_T(m) =& \frac{\int^w_{-w} \delta(x-m) \int^w_{-w} I_\text{in}(x,y) \,dy\,dx}{\int^w_{-w}\int^w_{-w} I_\text{in}(x,y) \,dy\,dx}
    \label{eq:lsfT_1}
\end{align}

Let us define the $y$-integral of the input beamlet intensity distribution as 

\begin{equation}
\mathcal{I}(x) = \int^w_{-w}I_\text{in}(x,y) \,dy.
\label{eq:yintegral}
\end{equation}

Inserting Equation \ref{eq:yintegral} into Equation \ref{eq:lsfT_1}, and applying the sifting property of the delta function, we show

\begin{align}
\begin{split}
    \text{LSF}_T(m) =& \frac{\int^w_{-w} \delta(x-m) \,\mathcal{I}(x)\,dx}{\int^w_{-w}\int^w_{-w} I_\text{in}(x,y) \,dy\,dx}
    \\
    =& \frac{ \mathcal{I}(m)}{\int^w_{-w}\int^w_{-w} I_\text{in}(x,y) \,dy\,dx}.
\end{split}
\end{align}

Given that $\int^w_{-w}\int^w_{-w} I_\text{in}(x,y) \,dy\,dx$ is a scalar independent of $m$, and the LSF is self-normalised (thus we can neglect constant scaling factors), we can prove

\begin{equation}
    \text{LSF}_T(m) = \mathcal{I}(m).
    \label{eq:lsfT}
\end{equation}

The LSF is thus defined on the object space, and is given by the projected input probe intensity distribution evaluated at the corresponding object space position. 
\par
We note that, where Equation \ref{FP_FD} holds, and for independent beamlets, the conserved Noether charge \cite{paganin2019x}

\begin{equation}
    \mathcal{N} = \int^w_{-w}\int^w_{-w} I(x,y,z) \,dx\,dy, \hspace{40pt} z \geq 0
    \label{eq:noether}
\end{equation}

demonstrates that phase and dark-field effects do not influence the measured transmission signal or its spatial resolution, and that the model would therefore apply also to more general objects.

\subsection{Refraction}

\label{sec:ref_deriv}

Consider a pure phase object such that

\begin{equation}
    I_z(x,y) = I_\text{in}(x,y)
    - \frac{z_\text{od}}{k} \nabla_\perp\cdot
    \left[
    I_\text{in}(x,y) \nabla_\perp\phi_0(x,y)
    \right].
    \label{eq:pure_phase_FP}
\end{equation}

For a system with 2D sensitivity, refraction is a vector quantity with components in both the $x$ and $y$ directions. In the moments model, the refraction angle in the $x$-direction is given by 

\begin{align}
\begin{split}
    \theta_x(m,n) &= \frac{M_{1x}[I_z(x,y)] - M_{1x}[I_\text{in}(x,y)]}{z_\text{od}}
    \\
    &= \frac{1}{z_\text{od}} \left[\frac{\int^w_{-w} x \int^w_{-w} I_z(x,y) \,dy\,dx}{ \int^w_{-w}\int^w_{-w} I_z(x,y) \,dy\,dx} - \frac{\int^w_{-w} x \int^w_{-w} I_\text{in}(x,y) \,dy\,dx}{ \int^w_{-w}\int^w_{-w} I_\text{in}(x,y) \,dy\,dx}   \right]. 
    \label{eq:refraction_moments}
\end{split}
\end{align}

Substituting Equation \ref{eq:pure_phase_FP} into Equation \ref{eq:refraction_moments}, and noting from the conserved Noether charge in Equation \ref{eq:noether} that for a non-attenuating object $\int^w_{-w}\int^w_{-w} I_\text{in}(x,y) \,dy\,dx =  \int^w_{-w}\int^w_{-w} I_z(x,y) \,dy\,dx $, we simplify to

\begin{align}
    \theta_x(m,n) 
    &=  -\frac{\int^w_{-w} x \int^w_{-w}   \nabla_\perp\cdot
    \left[
    I_\text{in}(x,y) \nabla_\perp\phi_0(x,y)
    \right] \,dy\,dx}{ \, k \int^w_{-w}\int^w_{-w} I_\text{in}(x,y)
    \,dy\,dx}. 
    \label{eq:theta_1}
\end{align}

For a refraction signal $\theta_x \propto \frac{\partial \phi}{\partial x}$, a Heaviside phase function generates a unit impulse response. Thus to analytically determine the refraction LSF we probe using a 1D Heaviside function $\phi(x,y) = H(x-m)$, with the property $ \nabla_\perp H(x - m) = \delta(x-m)$. Inserting the probe into Equation \ref{eq:theta_1}, neglecting constant scaling factors, and integrating by parts, we show

\begin{align}
\begin{split}
    \text{LSF}_\theta(m) &= -\int^w_{-w}\int^w_{-w} x \nabla_\perp\cdot
    \left[
    I_\text{in}(x,y) \, \nabla_\perp H(x-m)
    \right]
    \,dx \,dy \\
    &= -\int^w_{-w} x \frac{ \partial}{\partial x}
    \left[
    \mathcal{I}(x) \, \delta(x-m)
    \right]
    \,dx \\
    &= \int^w_{-w} \mathcal{I}(x) \delta(x-m) \,dx
    \\
    &= \mathcal{I}(m).
    \label{eq:lsftheta}
\end{split}
\end{align}

Thus proving that the LSF for the transmission and refraction channels are identical, for pure transmission and phase objects respectively.

\subsection{Dark-field}

Consider a pure phase object with inhomogeneities at a length scale below the probe size, such that it generates a diffuse scattering signal. We follow previous works in splitting the wave phase into fast $\phi_f$ and slow $\phi_s$ varying components \cite{yashiro2010origin,paganin2019x,esposito2023laboratory}, which contribute to the resolvable refraction signal $\theta_x \propto \phi_s'$ and the diffuse scattering angular variance $\alpha^2 \propto \text{var}(\phi_f')$, respectively. For a real phase-shifting object at X-ray energies ($\delta > 0, \,\therefore \, \phi <0$), a purely fast-varying phase with zero mean is not physically meaningful, thus the fast fluctuations must be supported by a slowly varying carrier phase $\phi_s < 0$. This provides justification to consider terms due to both resolvable phase and also diffuse scattering, particularly for dark-field signal close to edges, where $\phi_s(x)$ explicitly must vary quickly with respect to the illumination.  We thus describe the propagated beamlet intensity after interacting with the inhomogeneous phase object as 

\begin{equation}
    I_z(x,y) = I_\text{in}(x,y)
    -\frac{z_\text{od}}{k} \nabla_\perp \cdot \left[
    I_\text{in}(x,y)\nabla_\perp\phi_0(x,y)
    \right]
    + z_\text{od}\nabla_\perp^2 \left[ D_0(x,y) I_\text{in}(x,y) \right].
    \label{FP_DF}
\end{equation}

Again, choosing the $x$-component of the dark-field signal, and under the assumption of a Gaussian scattering distribution (a good approximation for ultra-small angle scattering \cite{lautizi2024universal}), we can retrieve the dark-field signal as the difference of variances

\begin{align}
    \alpha^2_x(m,n) &= \frac{ M_{2x}[I_z(x,y)] - M_{2x}[I_\text{in}(x,y)] - (M_{1x}[I_z(x,y)])^2}{z_\text{od}^2},
\end{align}

using $M_{1x}[I_\text{in}(x,y)] = 0$ after assuming $I_\text{in}(x,y)$ is a zero-centred even function. Equation \ref{FP_FD} is first order accurate in $z$, so we neglect the term $(M_{1x}[I_z(x,y)])^2$ which is second order in $z$. Again applying Noether conservation results in

\begin{align}    
\begin{split}
    \alpha^2_x(m,n) &= \frac{1}{z_\text{od}^2}\left[ 
    \frac{\int^w_{-w}x^2\int^w_{-w}I_z(x,y) \,dy\,dx}{\int^w_{-w}\int^w_{-w}I_z(x,y) \,dy\,dx  } - \frac{\int^w_{-w}x^2\int^w_{-w}I_\text{in}(x,y) \,dy\,dx}{\int^w_{-w}\int^w_{-w}I_\text{in}(x,y) \,dy\,dx}
    \right] \\
    &= \frac{\int^w_{-w}x^2\int^w_{-w} -\frac{1}{k} \nabla_\perp \cdot[I_\text{in}(x,y) \nabla_\perp \phi_0(x,y)] + \nabla_\perp^2 [I_\text{in}(x,y)D_0(x,y)] \,dy\,dx}{z_\text{od}\int^w_{-w}\int^w_{-w} I_\text{in}(x,y) \,dy\,dx}   .
\end{split}
\label{eq:darkfield_moments}
\end{align}

To analyse the unit input response to Equation \ref{eq:darkfield_moments}, we introduce scaled unit input functions $\phi_0(x,y) = \phi \delta (x-m)$ and $D_0(x,y) = D\delta(x-m)$, where $\phi$ and $D$ are scalars describing the relative strengths of the phase and diffuse scattering components. Beginning with the phase term, we integrate by parts and apply the sifting property of the delta function

\begin{align}
    \begin{split}
         \int^w_{-w}x^2\int^w_{-w} -\frac{1}{k}  \nabla_\perp \cdot[I_\text{in}(x,y) \nabla_\perp \phi_0 (x,y)] \,dy \,dx
        &= -\frac{\phi}{k} \int^w_{-w}x^2\frac{\partial}{\partial x} [\mathcal{I}(x) \frac{\partial \delta(x-m)}{\partial x}]  \,dx
        \\
        &= - \frac{2\phi}{k}  
     \left[ \mathcal{I}(m) + m \frac{\partial \mathcal{I}(m)}{\partial m} \right].
    \end{split}
\end{align}

For the diffuse scattering component, we similarly obtain

\begin{align}
    \begin{split}
    \int^w_{-w}x^2\nabla_\perp^2 [I_\text{in}(x,y)D_0(x,y)] \,dy\,dx =&  
    D \int^w_{-w}x^2  \frac{\partial^2}{\partial x^2} [\mathcal{I}(x)\delta (x-m)] \,dx
    \\
    =& 2D \, \mathcal{I}(m).
    \end{split}
\end{align}

\par

Combining terms, and neglecting constant scaling factors, we finally show that

\begin{align}
\begin{split}
    \text{LSF}_{\alpha^2}(m) =& \mathcal{I}(m) - \frac{\phi}{kD} \left[ \mathcal{I}(m) + m \frac{\partial \mathcal{I}(m)}{\partial m} \right].
    \label{eq:lsfDF}
\end{split}
\end{align}

The expression in Equation \ref{eq:lsfDF} indicates that the dark-field LSF is sample dependent, being affected by the relative strengths of the diffuse scattering and phase characteristics of the object. This arises from the mixing of phase and diffuse scattering signals near sample edges.

\subsection{Analytical optical transfer functions for circular and square apertures}

The exact shape of the input probe beamlet $I_\text{in}(x,y)$ is dependent not only on the shape of the aperture, but on its shape projected onto the sample plane. For a parallel beam with negligible source size, and the sample placed immediately after the modulator, $I_\text{in}(x,y)$ is well approximated by the true aperture shape. For a magnifying geometry, and finite source size, $I_\text{in}(x,y)$ is instead the magnified aperture, convolved with the projected source shape. For cases where the Fourier transform of the input probe $I_\text{in}(x,y)$ can be represented by an analytical function, it is now possible to calculate analytical OTFs to describe the contrast-specific spatial resolution. A similar process could be followed numerically when the probe is irregular or is convolved with a non-negligible projected source spot. 
\par
From Equations \ref{eq:lsfT}, \ref{eq:lsftheta}, and \ref{eq:lsfDF}, the transmission, refraction, and dark-field OTFs can be computed from the Fourier transforms of the LSFs as

\begin{align}
    \text{OTF}_{T\theta}(\omega) &= \mathcal{F}[\mathcal{I}(x)]
    \label{eq:otf_t_theta}
    \\
    \text{OTF}_{\alpha^2}(\omega) &=  \mathcal{F}[\mathcal{I}(x)] + \omega \frac{\phi}{kD} \frac{d}{d\omega} \mathcal{F}[\mathcal{I}(x)],
    \label{eq:otf_alpha}
\end{align}

where $\omega$ is the spatial frequency coordinate corresponding to the real-space $m$.

\par
For a circular aperture with radius R, the probe function and its $y$-integral may be described as

\begin{align}
    I_\text{circ}(x,y) =& 
    \begin{cases}
    0 & , \, \text{if } \sqrt{x^2+y^2} > R \\
   1 & , \, \text{if } \sqrt{x^2+y^2} \leq R
\end{cases},
\\
\mathcal{I}_\text{circ}(x) =& \, 2 \sqrt{R^2 - x^2}.
\label{eq:int_circ}
\end{align}

For a square aperture with width $L$,

\begin{align}
    I_{\text{rect}}(x,y) =& \begin{cases} 0 & , \, \text{if } |x| > L/2 \text{ or } |y| > L/2 \\ 1 & , \, \text{if } |x| \leq L/2 \text{ and } |y| \leq L/2 \end{cases}
    \\
    \mathcal{I}_\text{rect}(x) =& 
    \begin{cases} 
    0 & , \, \text{if } |x| > L/2 \\ 
    L & , \, \text{if } |x| \leq L/2 
    \end{cases},
    \label{eq:int_rect}
\end{align}

where up to constant scaling factors, Equation \ref{eq:int_rect} describes the $y$-integral of both a 2D square aperture or of a vertical 1D line aperture.
\par
Inserting Equations \ref{eq:int_circ} and \ref{eq:int_rect} into Equations \ref{eq:otf_t_theta} and \ref{eq:otf_alpha}, allows the calculation of the OTFs for transmission, refraction, and dark-field, for cases of 2D square and circular \cite{abramowitz1948handbook} apertures, or 1D line apertures. The resulting OTFs are summarised in Table \ref{tab:analyticalOTFs}.

\begin{table}[]
\renewcommand{\arraystretch}{1.5}
\centering
\begin{tabular}{|l|l|l|}
\hline
\textbf{Aperture} & \textbf{OTF}$_{T\theta}
(\omega)$ & \textbf{OTF}$_{\alpha^2}(\omega)$ \\ \hline
General form & $\mathcal{F}[\mathcal{I}(x)]$ & $\mathcal{F}[\mathcal{I}(x)] + \omega \frac{\phi}{kD} \frac{d}{d\omega} \mathcal{F}[\mathcal{I}(x)]$ \\ \hline
Circular & $\frac{J_1(2\pi \omega R)}{\pi \omega R}$  &  $ \frac{J_1(2\pi\omega R)}{\pi\omega R} - \frac{2 \phi}{\omega k D} J_2(2\pi\omega R)$ \\ \hline
          Square        & $\text{sinc}(\omega L)$  & $\text{sinc}(\omega L) + \frac{\phi}{kD}
        \left[ \text{cos}(\pi \omega L) - \text{sinc}(\omega L)\right]$ \\ \hline
\end{tabular}
\caption{Analytical optical transfer functions for apertures of arbitrary, circular, or square shape, based on the integrated probe shape $\mathcal{I}(x)$. The square aperture describes the OTF for both square and 1D line apertures, assuming that the direction of integration is parallel to the edges of the square. $J_1$ and $J_2$ are Bessel functions of the first and second kind respectively.}
\label{tab:analyticalOTFs}
\end{table}

\section{Experimental validation}

\subsection{Methods}

To verify the expressions derived in section \ref{sec:theory}, we compare to experimental results acquired using 2D beam-tracking with a synchrotron setup. A filtered (1.34 mm pyrolytic graphite, 2.1 mm aluminium) pink beam with a mean energy of 26.5 keV was obtained from the undulator at the Diamond Light Source I13-2 beamline. The beam was structured using a 100 \textmu{}m thickness tungsten modulator placed 221 m away from the source. The modulator consisted of laser-ablation fabricated 15 \textmu{}m diameter circular apertures, spaced in a regular square grid with a period of 50 \textmu{}m. The sample stage was located 150 mm downstream of the modulator, with the propagation distance $z_\text{od}$ set to 720 mm. Images were acquired using a pco.edge 5.5 camera coupled through magnifying optics to a scintillator screen, resulting in an effective pixel size of 2.6 \textmu{}m x 2.6 \textmu{}m, and a field-of-view of 6.7 mm x 5.6 mm. Considering the aperture width, the system with Fresnel number of 6.7 is thus within the near-field validity condition.
\par
To assess the system spatial resolution, a resolution target consisting of 1.5 \textmu{}m thickness gold on a 0.25 \textmu{}m thickness Si3N4 substrate was placed at the sample position. Projections were acquired at 32 x 32 dithering steps, obtained by shifting the modulator in steps of 1.5625 \textmu{}m in a 2D grid. At each position, two images of 300 ms each were acquired and averaged. Reference images were also acquired at each modulator position in the absence of the sample. Transmission, $x$-refraction, $y$-refraction, $x$-dark-field, and $y$-dark-field images were retrieved from the projections, through the moments-based method introduced in section \ref{sec:theory}. The phase signal was retrieved from the orthogonal refraction images through Fourier integration \cite{kottler2007two}.
\par
Additionally, to demonstrate that the same principles apply to low-coherence setups, a similar experiment was carried out using a laboratory-based system. An unfiltered polychromatic beam was obtained using the copper anode of the NXCT (National research facility for lab-based X-ray Computed Tomography) multi-contrast X-ray micro-CT system \cite{i2024new}, at source settings of 40 kV 30 mA. The beam was structured using a modulator placed 780 mm downstream of the source, consisting of 120 \textmu{}m of gold electroplated on a graphite substrate, with 1D apertures of 10 \textmu{}m width, and 79 \textmu{}m period. The sample was placed directly behind ($\sim$ 10 mm) the modulator, ensuring that the source of 70 \textmu{}m full-width-at-half-maximum (FWHM) has a negligible projected size ($\sim$0.9 \textmu{}m) through the aperture onto the sample plane. We note that this is much smaller than the $\sim$13 \textmu{}m projected source size on the detector plane, an additional 138 mm downstream, which does not have an impact on the beam-tracking spatial resolution. Images were acquired using a Hamamatsu Orca Flash4.0 V3 camera coupled through magnifying optics to a scintillator screen, resulting in an effective pixel size of 3.5 \textmu{}m x 3.5 \textmu{}m, and a field-of-view of 7.1 mm x 7.1 mm.
\par
Spatial resolution was assessed using a different resolution target, also consisting of 1.5 \textmu{}m thickness gold on a 0.25 \textmu{}m thickness Si3N4 substrate. Projections were acquired at 48 dithering steps, obtained by shifting the sample in steps of 1.667 \textmu{}m. At each position, five images of 6 s each were acquired and averaged. Reference images were also acquired in the absence of the sample. The transmission image was retrieved through the moments-based method, while $x$-refraction was retrieved by estimating shifts from beamlet-wise cross-correlation with Gaussian sub-pixel interpolation. 

\subsection{Results}

\begin{figure}[H]
    \centering
    \includegraphics[width=0.82\linewidth]{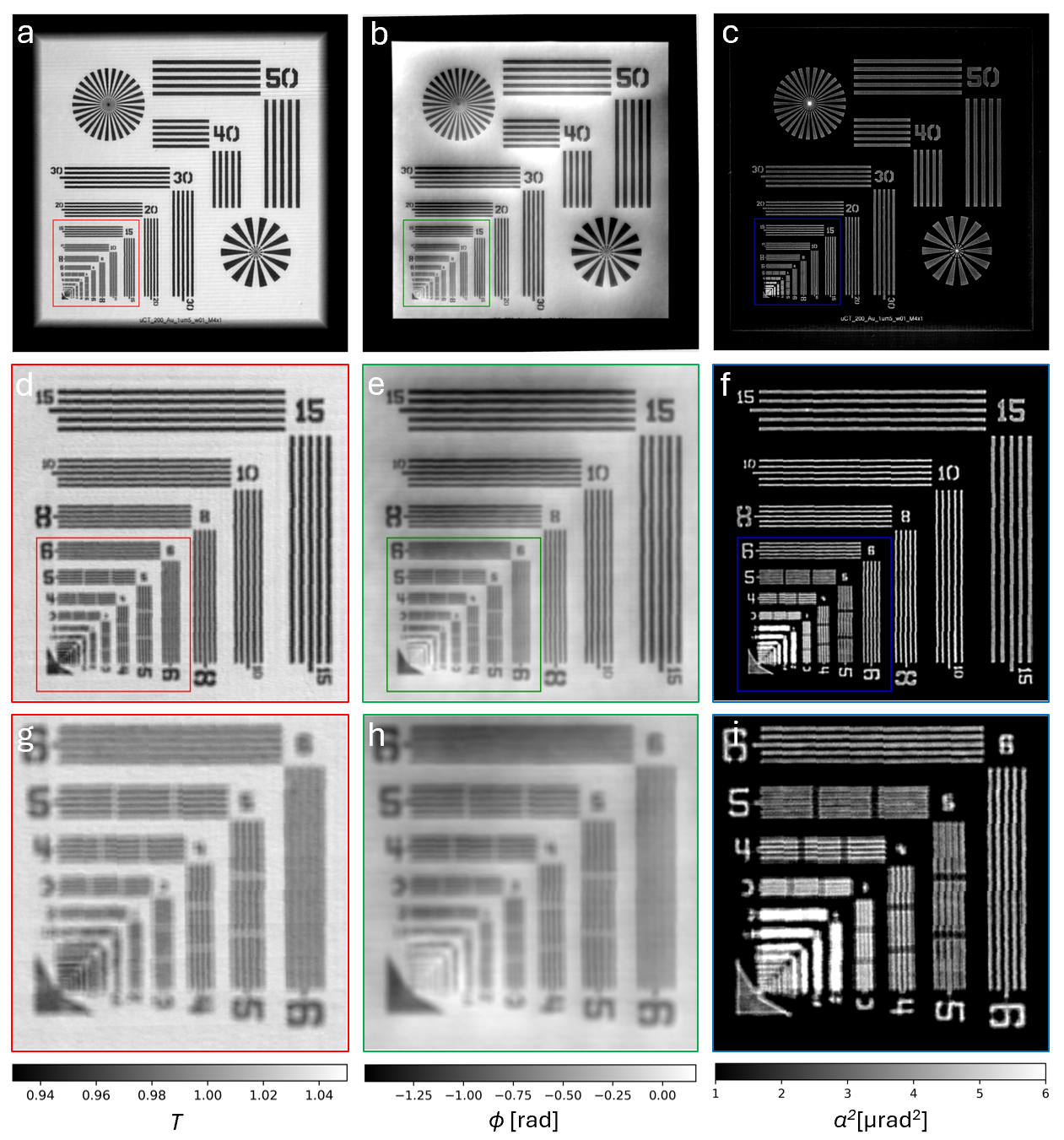}
    \caption{Retrieved transmission (a), phase (b), and dark-field (c) signals from the resolution target using a 2D beam-tracking synchrotron setup. Two further zoom levels are shown below each contrast channel, indicating the finest resolution bar patterns. In the transmission (g) and phase (h) contrast channels, the near zero contrast of the 6 \textmu{}m pattern corresponds closely to the expected zero crossing of the 15 \textmu{}m circular aperture. In the dark-field channel (i), contrast remains high at this same spatial frequency.}
    \label{fig:res_target}
\end{figure}

Figure \ref{fig:res_target} shows the transmission (\ref{fig:res_target}a), phase (\ref{fig:res_target}b), and dark-field (\ref{fig:res_target}c) signals retrieved from the synchrotron data, alongside close-ups of the finest resolution bar patterns. In each channel, the visibility of the pattern reaches a minimum, before increasing again with reversed-contrast, corresponding to zero-crossings in the OTF. For example, visibility drops to near zero for the 6 \textmu{}m bar pattern in transmission and phase, then returns for the 5 \textmu{}m pattern with a $\pi$ phase shift. This is in agreement with the first zero $[J_1(2\pi \omega R)]/(\pi \omega R) = 0$ at $\omega = 1.2197/2R$, which for the circular aperture of 15 \textmu{}m diameter results in $\omega = $ 81.31 lp/mm, or a real-space bar width of 6.15 \textmu{}m. Strong, non-reversed contrast is visible in the 5 \textmu{}m and 6 \textmu{}m patterns in the dark-field channel, beyond the conventional first zero of the aperture, giving a first indication of the different mechanism and increased sharpness in the dark-field channel.

\begin{figure}
    \centering
    \includegraphics[width=1\linewidth]{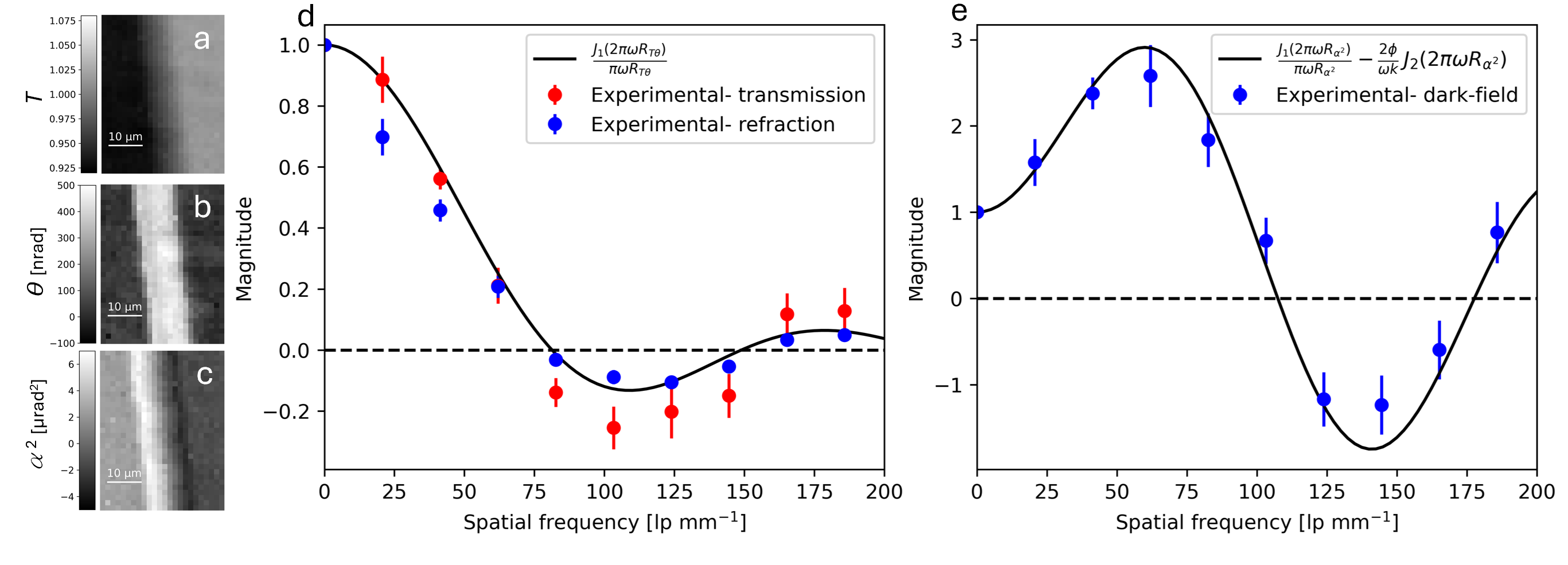}
    \caption{ROIs in the transmission (a), refraction (b), and dark-field (c) images used for calculating the OTFs for the synchrotron experiment. Experimental OTFs corresponding to transmission and refraction (a) and dark-field contrast channels (b). The solid black lines show the theoretical OTFs for a circular aperture of 15 \textmu{}m diameter. Negative OTF values indicate contrast reversal, which match the observations in Figure \ref{fig:res_target}. Note that d and e do not share the same $y$-axis.}
    \label{fig:OTF_experimental_theory}
\end{figure}

While the bar patterns give a strong indication of the resolving power of the system, the square waves contain additional odd harmonics in addition to the fundamental frequency of the bar \cite{coltman1954specification}. This adds a bias to the contrast measurements, which is further compounded by the high spatial frequency sensitivity of the derivative term in the dark-field OTF. Thus to quantitatively measure the frequency response of the system, we adopt the Fourier transform-based method approved by the International Organization for Standardization (ISO) \cite{ISO12233:2024}. A region-of-interest (ROI) was selected straddling a vertical edge on the larger Siemens star on the calibration standard, shown in Figure \ref{fig:OTF_experimental_theory} a, b, and c for the transmission, refraction, and dark-field images, respectively. The modulator step size of 1.5625 \textmu{}m is far beyond the Nyquist limit for the expected LSF from the 15 \textmu{}m aperture. The row-wise edge spread function (ESF) was calculated for the transmission and dark-field ROIs, while the refraction ROI was left as is due to already being a differential signal (see Section \ref{sec:ref_deriv}). The modulation transfer function (MTF) was calculated from the average of the row-wise Fourier transforms of the ESFs, with the standard deviation used to generate error bars. As the MTF captures the absolute value of the OTF, values were then multiplied by the sign of their phase based on visual observation of Figure \ref{fig:res_target}, to yield the OTFs for each contrast channel.
\par
Figure \ref{fig:OTF_experimental_theory} shows the experimental OTFs, alongside the corresponding theoretical OTFs. The theoretical OTFs were produced using the circular aperture models from Table \ref{tab:analyticalOTFs}, with the scalar constants $\phi = -0.66$ rad and $D = \frac{1}{2} \alpha^2z_\text{od} = 1.8 \times 10^{-12}$ rad$^2$m being extracted directly from the experimental measurements of the phase and dark-field images respectively. The model accurately captures the behaviour of the OTFs of all three contrast channels, including in predicting the location of zero crossings. The first-zeroes of the OTFs are at 81 lp mm$^{-1}$ for transmission and phase, and 107 lp mm$^{-1}$ for dark-field, confirming the greater spatial resolution in the dark-field channel. The shifting of the OTF peak away from zero frequency is analogous to the sample-dependent shift of the MTF peak observed in propagation-based phase imaging (PBI) \cite{sun2017mtf,ghani2019impact}. The phase part of the contrast transfer function \cite{pogany1997contrast} in PBI boosts high spatial-frequencies in the mixed attenuation-phase image, the same as it does in the beam-tracking dark-field image.

\par
\begin{figure}
    \centering
    \includegraphics[width=0.9\linewidth]{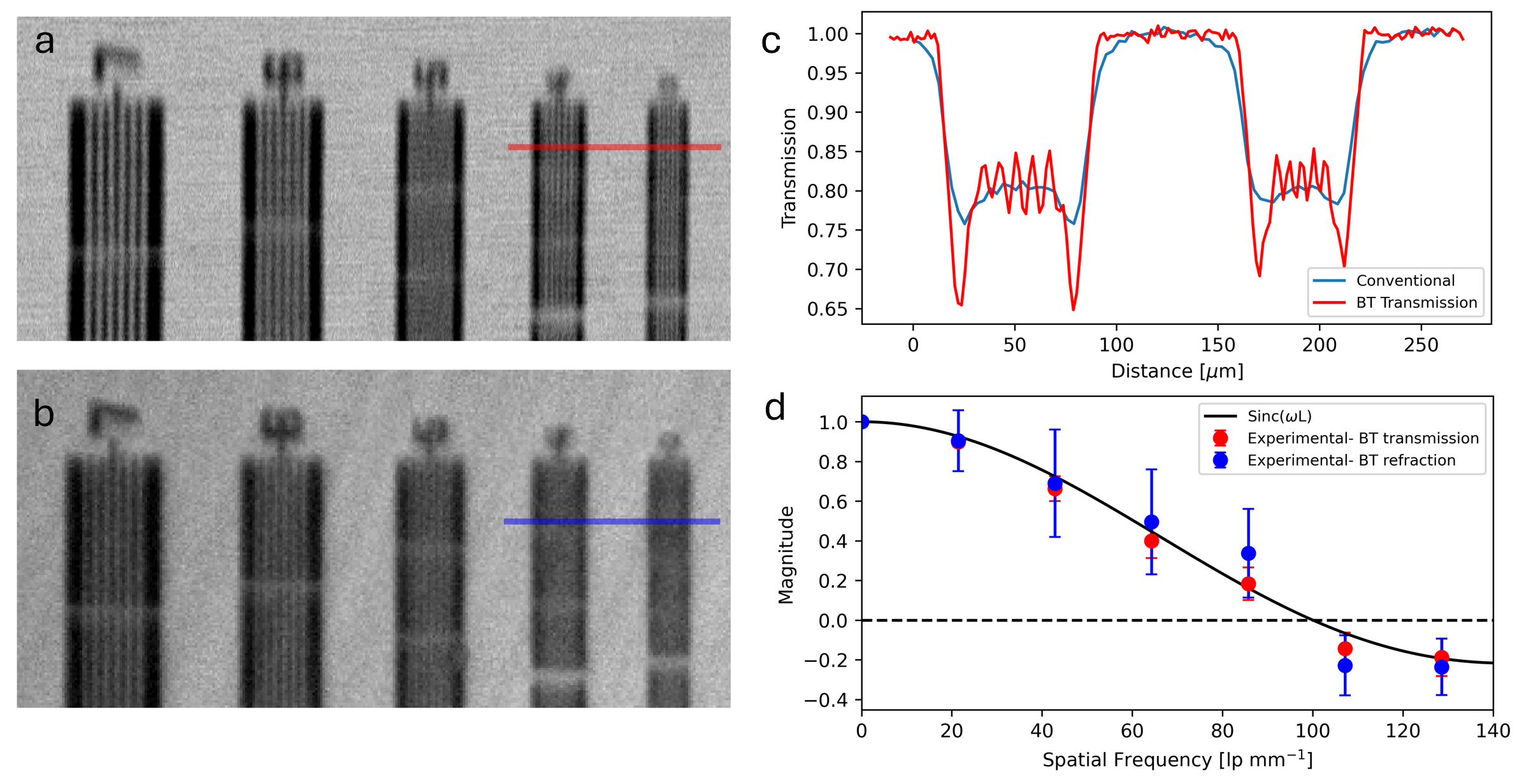}
    \caption{The retrieved lab-based beam-tracking transmission image (a) is shown in contrast to the equivalent conventional image obtained in the absence of the modulator. Line profiles (c) demonstrate that the beam-tracking image retains modulation down to the 4 \textmu{}m and 3 \textmu{}m half-width patterns, beyond the resolution limit of the conventional setup. The experimental OTFs for the beam-tracking transmission and refraction images are consistent with the theoretical OTF from the derived model.}
    \label{fig:lab_BT_fsp_figure}
\end{figure}

Figure \ref{fig:lab_BT_fsp_figure}a shows a section of the retrieved transmission image from the lab setup, focusing on line patterns of 7 \textmu{}m down to 3 \textmu{}m half-width. This is contrasted with the equivalent conventional transmission image obtained without the modulator, which is shown in Figure \ref{fig:lab_BT_fsp_figure}b. The line profile shown in Figure \ref{fig:lab_BT_fsp_figure}c demonstrates the presence of modulation on the 4 \textmu{}m and 3 \textmu{}m patterns in the beam-tracking image, where no modulation is present for the conventional image, indicating the ability of beam-tracking to surpass the resolution limits imposed by the source and detector. The lack of modulation on the 5 \textmu{}m pattern in Figure \ref{fig:lab_BT_fsp_figure}a corresponds to the first-zero in the OTF, which occurs at 100 lp/mm for the 10 \textmu{}m aperture. Figure \ref{fig:lab_BT_fsp_figure}d shows experimental OTFs obtained from a sharp edge in the transmission and refraction images, which are consistent with the theoretical OTFs from Table \ref{tab:analyticalOTFs}. Due to the lower brilliance of the laboratory source, and the weak scattering of the resolution target, the contrast-to-noise ratio of the retrieved dark-field image was not sufficient for analysis.

\section{Discussion}

We have developed an analytical model for spatial resolution in X-ray beam-tracking, allowing a description of the expected optical transfer functions for each contrast channel. Our results indicate both similarities and differences to the edge-illumination spatial resolution model \cite{diemoz2014spatial}, to which beam-tracking is often compared. The results are most  comparable in the transmission channel, where the spatial resolution is defined in both cases by the shape of the sampling illumination. In the beam-tracking case, this is given directly by the shape of the probe at the sample position, while in edge-illumination it is additionally sampled by a detector mask, which can narrow the contributing illumination further. The greater difference comes when considering phase effects. Unlike beam-tracking in which the entire beamlet is resolved and sampled at once (indicating that single-mask edge-illumination variants \cite{krejci2010hard,kallon2017comparing,shah2025application} would be best described using the beam-tracking model), in edge-illumination the beamlet is sampled by varying the offset between the sample and detector masks. As a consequence, the signal recorded can depend also on the width of the Fresnel fringe, which may induce a signal even when the sample is not aligned with the detector mask aperture (see Figure 4a in \cite{diemoz2014spatial}). While no expression for dark-field spatial resolution in edge-illumination has been derived previously, the similar sensitivity of the method to both resolvable phase effects and diffuse scattering suggests that a similar improvement in spatial resolution should also be present. This is supported by empirical observations in other works \cite{doherty2023edge,doherty2024hybrid}.
\par
Our model and experimental results have examined an idealised model of beam-tracking, in which beamlets are entirely independent. In many applications though, stepping requirements and scan times are kept short by using modulators with a smaller period relative to the detector resolution, which introduces cross-talk between beamlets. Similar to the correlated signal in edge-illumination \cite{havariyoun2023modulation}, this introduces an anti-correlated signal between adjacent beamlets (Appendix 1). While this introduces minor side lobes to the LSF, the shape of the central lobe remains unchanged, and could be managed using retrieval methods that consider correlation with adjacent beamlets \cite{jones2018retrieval}. Future work modelling beamlet crosstalk could aid modulator design, to maximise flux while minimising crosstalk.
\par
Applications of the derived model include experimental and system design. For example, for a desired spatial-resolution at 10\% contrast, it is possible to calculate the required square aperture width from the transcendental equation sinc($\omega L$) = 0.1, leading to $\omega L \approx 0.91$. For a target resolution of 1 \textmu{}m in the transmission or phase channels, the maximum permittable aperture width would thus be 1.8 \textmu{}m. Similarly, given the resolution at 10\% modulation, one could use this to decide the sampling step size in dithered acquisitions. The derived OTFs can additionally be used to partially recover information past the first zero of the aperture, providing justification to increase the sampling rate beyond the Nyquist frequency. Direct Fourier inversion using only the sign of the OTF recovers the phase of these frequencies without amplifying magnitudes, and therefore not introducing the numerical instabilities associated with deconvolution (Appendix 2).

\section{Conclusion}

We have derived a model describing the contrast-specific OTFs in X-ray beam-tracking microscopy, which has been confirmed by comparison to experimental results. A key finding of the model is that the dark-field channel is capable of resolving finer features than either transmission or phase, in part due to its sensitivity to changes in phase at sample edges. In all cases, features much smaller than the aperture diameter are resolvable. The model applies to both synchrotron and lab-based setups, and demonstrates that beam-tracking can push spatial-resolution beyond the limits imposed by the source and detector. Future work will investigate further the dark-field signal around edges, and the possible applications of the improved spatial resolution that it offers.

\section{Acknowledgements}

This work is supported by the EPSRC-funded UCL Centre for Doctoral Training in Intelligent,
Integrated Imaging in Healthcare (i4health) (EP/S021930/1), the Department of Health’s
NIHR funded Biomedical Research Centre at University College London Hospitals, and the
National Research Facility for Lab X-ray CT (NXCT) through EPSRC grants EP/T02593X/1
and EP/V035932/1; and by the Wellcome Trust 221367/Z/20/Z. We gratefully acknowledge Diamond Light Source for time on beamline I13-2 under proposal MG37963-1.

\bibliography{refs}



\newpage

\section*{Appendix}

\subsection*{1 Detector crosstalk}

\begin{figure}[h]
    \centering
    \includegraphics[width=0.85\linewidth]{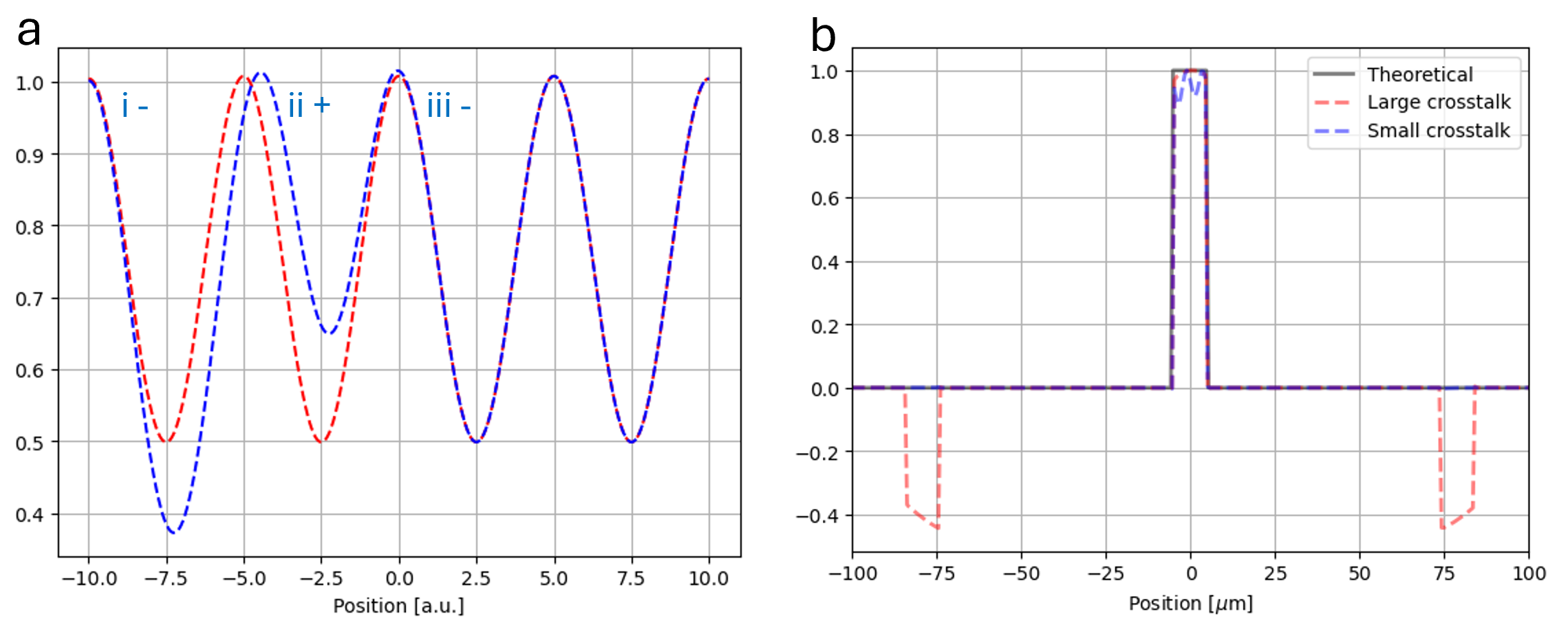}
    \caption{Beamlets which are not adequately separated in relation to the system resolution can experience crosstalk effects (a). If the beamlet ii experiences a positive shift, crosstalk causes an increase in counts on the negative tail of its neighbouring beamlet iii, thus inducing a small negative signal. Similarly, this causes a reduction in counts on the positive tail of beamlet i, also inducing a small negative signal. The theoretical line spread function of a 10 \textmu{}m width 1D aperture (b) is compared to simulated results in the presence of small (5 \textmu{}m FWHM) and large (80 \textmu{}m FWHM) amounts of detector crosstalk. While the size of the main body of the line spread function does not change, large crosstalk introduces secondary lobes corresponding to signal in the adjacent apertures.}
    \label{fig:crosstalk}
\end{figure}

The derived model has assumed an ideal imaging system, in which we have assumed negligible crosstalk between beamlets. Translation of beam-tracking to the laboratory often relies on modulators with a smaller period relative to the detector resolution, in order to reduce stepping requirements and scan times. Reducing modulator periods, or decreasing the spatial resolution of the system causes the signal in one beamlet to be correlated with the signal in its neighbours. This is demonstrated in Figure \ref{fig:crosstalk}a, where it is shown how a positive refraction signal in beamlet ii induces a small negative signal in beamlets i and iii. 

The effect of detector crosstalk on the LSF was simulated using a verified wave optics code \cite{vittoria2013strategies}. We generate a refraction signal from a 10 \textmu{}m thickness low-density-polyethylene (LDPE) edge. The simulated system featured a monochromatic 18 keV 1 \textmu{}m full-width-at-half-max (FWHM) source, 50 m upstream of a 1D modulator, consisting of 10 \textmu{}m width apertures with 79 \textmu{}m period. The LDPE edge was placed at the modulator plane, aligned parallel to the apertures. The detector, 10 mm downstream of the sample, consisted of 5 \textmu{}m pixels, with an LSF of 5 \textmu{}m in the small crosstalk case, and an LSF of 80 \textmu{}m in the large crosstalk case. A series of 128 dithering steps across one period were simulated, followed by moments-based retrieval. The refraction signals are compared in Figure \ref{fig:crosstalk}b to the theoretical LSF derived from Equation \ref{eq:lsftheta}. In the small crosstalk case, an excellent agreement with the theoretical model is found. In the large crosstalk case, the central lobe of the LSF still matches the theoretical probe width, demonstrating the ability of beam-tracking to far exceed detector limited resolution. The crosstalk does however give rise to secondary lobes, corresponding to adjacent beamlets and thus separated by one period from the central lobe, similar to previous observations in edge-illumination \cite{havariyoun2023modulation}. We note however, that a particularly large LSF, much larger than typical for similar experimental setups, was chosen in order to exaggerate crosstalk effects. While a moments-based retrieval method has been adopted throughout this work, methods such as multi-Gaussian fitting \cite{jones2018retrieval} which explicitly account for crosstalk may reduce the prominence of secondary lobes.

\subsection*{2 Recovering reversed contrast}

\begin{figure}[h]
    \centering
    \includegraphics[width=0.7\linewidth]{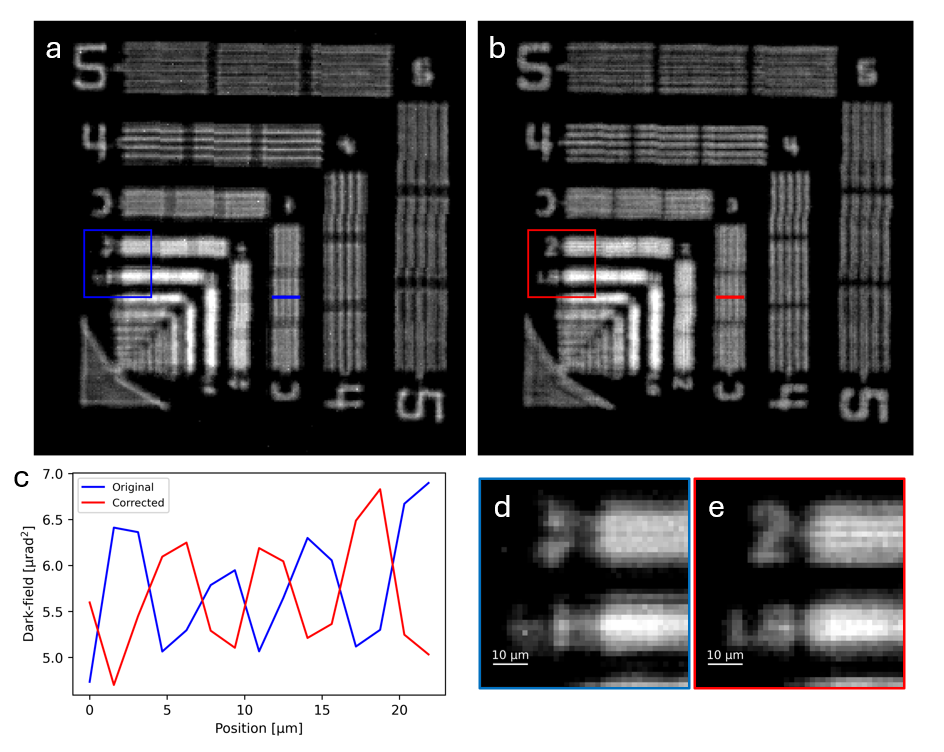}
    \caption{Direct Fourier inversion of the contrast reversal in the dark-field image (a) results in the image (b) in which contrast reversal is corrected without amplifying high frequencies. Line profiles through the 3 \textmu{}m pattern are shown in (c), illustrating the return of the correct phase of the pattern. Despite no attempt to deconvolve higher frequencies, correcting the contrast reversal makes the blurred numbers (d) legible (e), and recovers some contrast even on the 2 \textmu{}m bar pattern.}
    \label{fig:deconvolved_reversal}
\end{figure}

Figure \ref{fig:OTF_experimental_theory} demonstrates the presence of contrast far beyond the first zero of the OTF, especially for the dark-field channel. Past this point, contrast reversal can cause destructive interference and loss of information. Some contrast however can be restored through Fourier inversion by 

\begin{equation}
    I_\text{out}(x,y) = \mathcal{F}^{-1} \left[ \frac{\mathcal{F}[I_\text{blur}(x,y)]}{\text{sng}\{\text{OTF}(u,v)\}}  \right],
    \label{eq:deconv_sign}
\end{equation}

where $I_\text{blur}(x,y)$ is the input image, and $I_\text{out}(x,y)$ is the image after correcting for contrast reversal. The signum function sng returns the sign of its argument, thus capturing only the phase, and not the magnitude, of the OTF. The consequent lack of division close to zero means that high frequency noise is not amplified, and unlike true deconvolution regularisation is not required. Figure \ref{fig:deconvolved_reversal} demonstrates the application of Equation \ref{eq:deconv_sign} on the dark-field image of the resolution target. Correcting the reversed phases shown in Figure \ref{fig:deconvolved_reversal}c recovers the structure of complex features such as the numbers "2" and "1.5" in \ref{fig:deconvolved_reversal}e, and even recovers some contrast on the 2 \textmu{}m bar pattern. Note that it is also similarly possible to use the full OTF for conventional deconvolution, using regularisation methods such as Wiener filtering.

\end{document}